\documentclass[twocolumn]{aastex7}

\newcommand\teff{$T_{\mathrm{eff}}$}
\newcommand\logg{$\log{g}$}
\newcommand\vsini{$v\sin{i}$}
\newcommand\vmac{$v_{\mathrm{mac}}$}
\newcommand\vmic{$v_{\mathrm{mic}}$}
\newcommand\vrad{$v_{\mathrm{rad}}$}

\usepackage{xcolor}

\shorttitle{AASTeX v6.3.1}
\shortauthors{Claire Komori}

\graphicspath{{./}{figures/}}

\begin{document}

\title{The Effects of Sunspots on Spectral Line Shapes in the Visible}

\author[0009-0005-9362-6778]{Claire Komori}
\affiliation{Department of Physics and Astronomy, San Francisco State University, 1600 Holloway Ave, San Francisco, CA 94132, USA}
\email{ckomori@sfsu.edu}

\author[0000-0002-9873-1471]{John M. Brewer}
\affiliation{Department of Physics and Astronomy, San Francisco State University, 1600 Holloway Ave, San Francisco, CA 94132, USA}
\email{jmbrewer@sfsu.edu}

\author[0000-0002-3852-3590]{Lily L.\ Zhao}
\thanks{NASA Sagan Fellow} 
\affil{Center for Computational Astrophysics, Flatiron Institute, 162 Fifth Avenue, New York, NY 10010, USA}
\affil{Department of Astronomy \& Astrophysics, University of Chicago, 5640 S Ellis Ave, Chicago, IL 60637, USA}
\email{lilylingzhao@uchicago.edu}

\begin{abstract}
We present a comparative spectral analysis to explore the ability of a cooler Sun model to accurately capture the spectral line shape changes caused by Sunspots. In the search for small Earth-like planets, the effects of stellar surface activity can overwhelm the $\sim$10 cm/s planetary RV signal. This necessitates the development of new stellar modeling methods and a greater understanding of the impact of surface activity on stellar spectra. Some attempts to model out noise from Sunspot activity, in particular, have used a sum of a stellar model with a model of a cooler, but otherwise identical star. From our analysis, we find that a cooler effective temperature alone cannot capture the numerous spectral line shape variations seen in a Sunspot observation. The cooler temperature of a Sunspot not only deepens the cores of atomic lines, it also increases and strengthens molecular lines that are not fully represented in our line list. Furthermore, our LTE models and a comparison cool star also fail at capturing line strengthening, broadening, blending, and splitting induced by the magnetic field in the Sunspot. 

\end{abstract}

\section{Introduction} \label{sec:intro}

The radial velocity (RV) method is one of the primary techniques for exoplanet detection, used in identifying over 1000 planets thus far \citep{Mayor_1995, Lee_2023, Li_2024}. In the last few years a number of extreme-precision radial velocity (EPRV) instruments such as EXPRES \citep{Petersburg_EXPRES_2020}, NEID \citep{Gupta_NEID_2021}, ESPRESSO \citep{Pepe_ESPRESSO_2021}, and MAROON-X \citep{Seifahrt_MAROONX_2022}, have pushed down to $\sim$30 cm/s precision, leading to many high resolution, high signal-to-noise time series spectra, and allowing for the discovery of lower-mass planets at longer orbital periods \citep{Faria_2022, Basant_2025}. Additionally, these spectra can be used to derive important information regarding the host-star itself such as its elemental abundance, surface gravity, magnetic field, and stellar activity. Correctly interpreting stellar spectra is crucial for characterizing host-stars and furthering planet detection. 

RV measurements are sensitive to not just the Doppler shift from reflex motions induced by the companion planet(s), but also to signals induced by the star itself \citep{Dumusque_2016, SiuTapia_2017, Zhao_2022}. At the $\sim$30 cm/s stability typical of EPRV instruments, stellar surface variability is now the dominant source of error veiling potential low-mass planetary signals. The semi-major amplitude of the reflex velocity of the Sun due to the Earth in its one year orbit is only $\sim$9.5 cm/s. Low-mass planets, and even sub-Neptunes on longer period orbits, are hidden by stellar surface variability requiring the mitigation of the stellar signals to below 10 cm/s. One origin for stellar variability are starspots. \cite{Saar_Donahue_1997}, in particular, showed that perturbations caused by Sunspots can increase with stellar rotation and level of activity. The combination of the star's rotation with starspots can mimic periodic RV signals and potentially lead to false positive planet detection \citep{Boisse_2011, Vanderburg_2016, Robertson_2020}, and these misinterpretations have occurred in the past in cases such as TW Hydrae \citep{Huelamo_2008} and LkCa19 \citep{Huerta_2008}. Surface features on the star may alter line shapes across the spectrum in a variety of ways. As we attempt to extract ever more precise RVs from these spectral lines, accurate knowledge of how individual lines are changing is needed.

Activity indicators are often used to disentangle stellar signals, such as Sunspots, from RV measurements. Spectral line features such as emission in the core of Ca II H\&K lines (3969.6{\AA} and 3934.7{\AA} respectively; \citet{Saar_1998, Meunier_Lagrange_2013}), the Ca infrared triplet (8498{\AA}, 8542{\AA}, and 8662{\AA} respectively; \cite{Saar_Fischer_2000}), and the H$\alpha$ line (6562.8{\AA}; \cite{Skelly_2008, Robertson_2014, Giguere_2016}) are used to gauge the presence and level of magnetic activity \citep{vonStauffenberg_2024}. Attributes of a cross-correlation function (CCF) between the spectra and a binary line list are also often used.  These attributes include CCF bisector asymmetry measurements and the full width half maximum (FWHM) of the CCF, which are used to trace asymmetrical line shape changes that are seen in most lines throughout the observed spectrum \citep{Lafarga_2020, Lafarga_2023}. However, these activity indicators are sensitive only to global changes or changes in specific spectral regions.

There have been further developments regarding the use of activity indicators and methods of disentangling stellar variability signals from planetary RV signals in recent times. The use of the Gaussian processes (GPs) have allowed for the modeling of quasi-periodic signals that are potentially driven by stellar surface activity \citep{Jones_2017, Gilbertson_2020, Rajpaul_2020}. GPs are used to construct flexible stellar activity models that trace RV signals that are also reflected in the time series of activity indicators. However, this will only be able to model stellar signals captured by the activity indicators. Additionally, the flexibility of such models may allow the activity model overfit the data, potentially fitting out small planet signals.

A comparative test of different stellar activity modeling and mitigation methods showed a lack of agreement between returned planetary RVs after being cleaned of stellar activity signals \citep{Zhao_2022}, which necessitates continued progress in this area. Furthermore, the spectroscopic data of stellar activity signals in itself can be a significant source of rich information regarding a star’s magnetic field which generates the photospheric and chromospheric activity.

Local thermodynamic equilibrium (LTE) models are often used for interpreting integrated light spectra. However, it faces significant challenges when stars exhibit surface variation. Typical LTE spectral models \citep[e.g.][]{Przybilla_2011} are based on the assumption of a quiet star; they lack the ability to account for line shape changes caused by stellar activity. This can lead to incorrect inference of fundamental parameters such as temperature, surface gravity, and elemental abundances. 

Here, we investigate line shape changes seen in a Sunspot spectrum as a step towards improving spectral modeling methods. Sunspots are magnetic structures that appear dark on the Solar surface \citep{Solanki_2003}. The inner darker regions called the umbra are 1000-1900K cooler than the quiet Solar surface \citep{Solanki_2003}. Sunspots appear on points where there are strong magnetic fields. In these areas, the magnetic force suppresses convection from below the photosphere, making it difficult for the hot gas to rise \citep{Thomas_1992}. This results in these areas having cooler temperatures in comparison to the rest of the Solar surface and appearing as dark Sunspots. 

One strategy used in an attempt to model stars with Spot activity is by summing two stellar spectra with different temperatures for a given Spot filling factor, Spot contrast, and observed \teff\ \citep{Morris_2019, Cao_2022, Wilson_2023}. This is done by combining a stellar template at some ambient temperature with that of a template of a cooler, but otherwise identical, star; by construction this yields the observed effective temperature at the observed flux. However, this is a simplification of a complex problem, as there is yet to be a spectral modeling method that is able to incorporate the various non local thermodynamic equilibrium (NLTE) and magnetic field effects that can occur with spotted stars. For example, line strengths in Sunspots have been shown to increase for some atomic species and molecules while weakening for others \citep[][and references therein]{Wallace_2005}. Some of this is due to chemical changes (i.e. atoms being incorporated into molecules). Simultaneous modeling of both atomic and molecular species would be needed to account for such changes. 
 Magnetic effects on some lines adds an additional challenge.

There have been many studies of individual spectral lines in Solar and Sunspot spectra. The magnetically sensitive neutral iron line has been especially well studied since the late 1900s, with some focusing on the NLTE line effects \citep{Rutten_1982, Solanki_1988, Rutten_1988}, while others focused particularly on photospheric magnetism \citep{Horn_1997, Lites_1998, Balthasar_1999}, but overall photospheric iron lines have been generally modelled assuming LTE. There is continued investigation on the role of NLTE effects on the formation of iron lines in the Solar spectra \citep{Smitha_2020, Smitha_2021, Smitha_2023}, while others have analyzed individual lines to measure magnetic field strengths of Sunspots \citep{Lozitsky_2016, Lozitsky_2020, Kuckein_2021, Mathur_2024}. 

In particular, \citet{Smitha_2020, Smitha_2021} focused on the 6301-6302{\AA} iron line pair and also studied another widely utilized iron line at 6173{\AA} in \cite{Smitha_2023}. These studies looked into the dominant NLTE mechanism that induce Zeeman splitting, and investigated the influence of neglecting NLTE conditions on the inference of stellar parameters. The 6301-6302{\AA} iron line pair is affected by UV irradiation, which causes the overionization of iron atoms and leads to deviations from LTE conditions. 3D magnetohydrodynamic simulations were used to compute the Stokes profiles of the iron lines which were then inverted into LTE and NLTE to infer the atmosphere. It was found that neglecting NLTE effects around magnetic structures can cause errors in derived temperature \citep{Smitha_2020}, as well as, to a lesser extent, other atmospheric parameters such as line-of-site velocity and magnetic field strength \citep{Smitha_2021}. 

In our work, we use a Sunspot umbral observation spanning from 5162{\AA} to 6664{\AA} \citep{Wallace_2005}, to explore the ability of a cooler Sun model to accurately reproduce the features found in a Sunspot spectrum. We also construct a best-fit model to derive stellar parameters of a Sunspot spectrum. By comparing the observed spectrum against model spectra, we investigate the different line shape variations seen in the Sunspot observation. Additionally, we conduct empirical comparisons between the Sunspot observation and an observation of a cool star with nearly Solar parameters, but with an effective temperature of that of our best-fit model. This comparison therefore incorporates any NLTE differences and molecular species at that lower temperature that may be lacking in our model.

\section{Observation} \label{sec:obs}
We use a high-resolution spectrum of a Sunspot umbra covering part of the visible spectrum taken with the Fourier Transform Spectrograph (FTS) on the McMath-Pierce Solar Telescope at Kitt Peak Observatory. This observation of the umbra was originally made by L.\ Testerman, recorded as 1981/03/25\#1, and re-reduced in 2000 and 2005 by \cite{Wallace_2005}. The data is available  as \dataset[sunspot atlas 4]{https://nispdata.nso.edu/ftp/pub/atlas/spot4atl/} at the NSO archives. The entire spectrum, which also includes contributions  from other observations, spans between 3920{\AA} to 6664{\AA}.

The now decommissioned McMath-Pierce Solar Telescope was designed to study the structure and spectra of Sunspots. The telescope operated FTS had typical resolutions between 350,000 and 700,000 and S/N ratio that ``exceeds many hundreds'' \citep{Wallace_2011}. The FTS works by reconstructing a spectrum from the interference pattern through Fourier transformation \citep{saptari2003fourier}. The observations are continuum normalized and telluric lines were divided out \citep{Wallace_2005}. The available data do not include uncertainties.

To make an empirical comparison of the Sunspot spectrum against a cool star, we chose one of the spectra of a star from \citet{Brewer_2016} that closely matched the Sun in surface gravity \textbf{(\logg)} and metallicity \textbf{([M/H])}, but had an effective temperature \textbf{(\teff)} close to that of our best-fit spectral model of the Sunspot. The chosen target, HD 130992, is one out of the 1617 stars that were observed by the HIRES spectrograph at Keck Observatory \citep{Vogt_1994} as part of the California Planet Survey \textbf{(CPS; \cite{Howard_2010})}, for which \citet{Brewer_2016} conducted a uniform spectroscopic analysis to obtain stellar parameters. HD 130992 is a cool dwarf with \teff=4767K, \logg=4.51, [M/H]=-0.03, \vsini=0.7 km/s, \emph{v}\textsubscript{mac}=1.7 km/s, and R'\textsubscript{HK} (activity index)=-4.75 \citep{Brewer_2016}. These parameters were derived with an LTE model. However, affects arising from the cooler temperature or activity that are not captured by the model should be visible in the cool star observation. The observed spectrum of HD 130992 also covers a larger spectral range than our model (see Section \ref{sec:sme}).

\begin{deluxetable*}{cccccccccc}
\tablecaption{Model Parameters}
\tablenum{1}

\tablehead{\colhead{Model} & \colhead{Temperature} & \colhead{log \emph{g}} & \colhead{[M/H]} & \colhead{\emph{v} sin\emph{i}} & \colhead{\emph{v}\textsubscript{mac}} & \colhead{\emph{v}\textsubscript{mic}} \\ 
\colhead{} & \colhead{(K)} & \colhead{} & \colhead{} & \colhead{(km/s)} & \colhead{(km/s)} & \colhead{(km/s)} & \colhead{}} 

\startdata
Solar & 5777 & 4.44 & 0.00 & 1.63 & 3.5 & 0.85 \\
cooler & 5170 $\pm$ 24 & 4.44 & 0.00 & 1.63 & 3.5 & 0.85 \\
best-fit & 4767 $\pm$ 24 & 3.68 $\pm$ 0.05 & -0.418 $\pm$ 0.03 & 0.0$^*$ & 7.15$^*$ $\pm$ 0.5  & 0.85 \\
\enddata
\tablecomments{\label{table:1}The stellar parameters for the three models (Solar, cooler, and best-fit) constructed by forward modeling using SME. {$^*$} Best-fit value for \vmac\ derived with \vsini\ fixed to zero. A typical value for the integrated disk quiet Sun when fit this way is $< 4.0$ km/s.}
\end{deluxetable*}

\begin{figure*}[htb!]
\includegraphics[width=\linewidth]{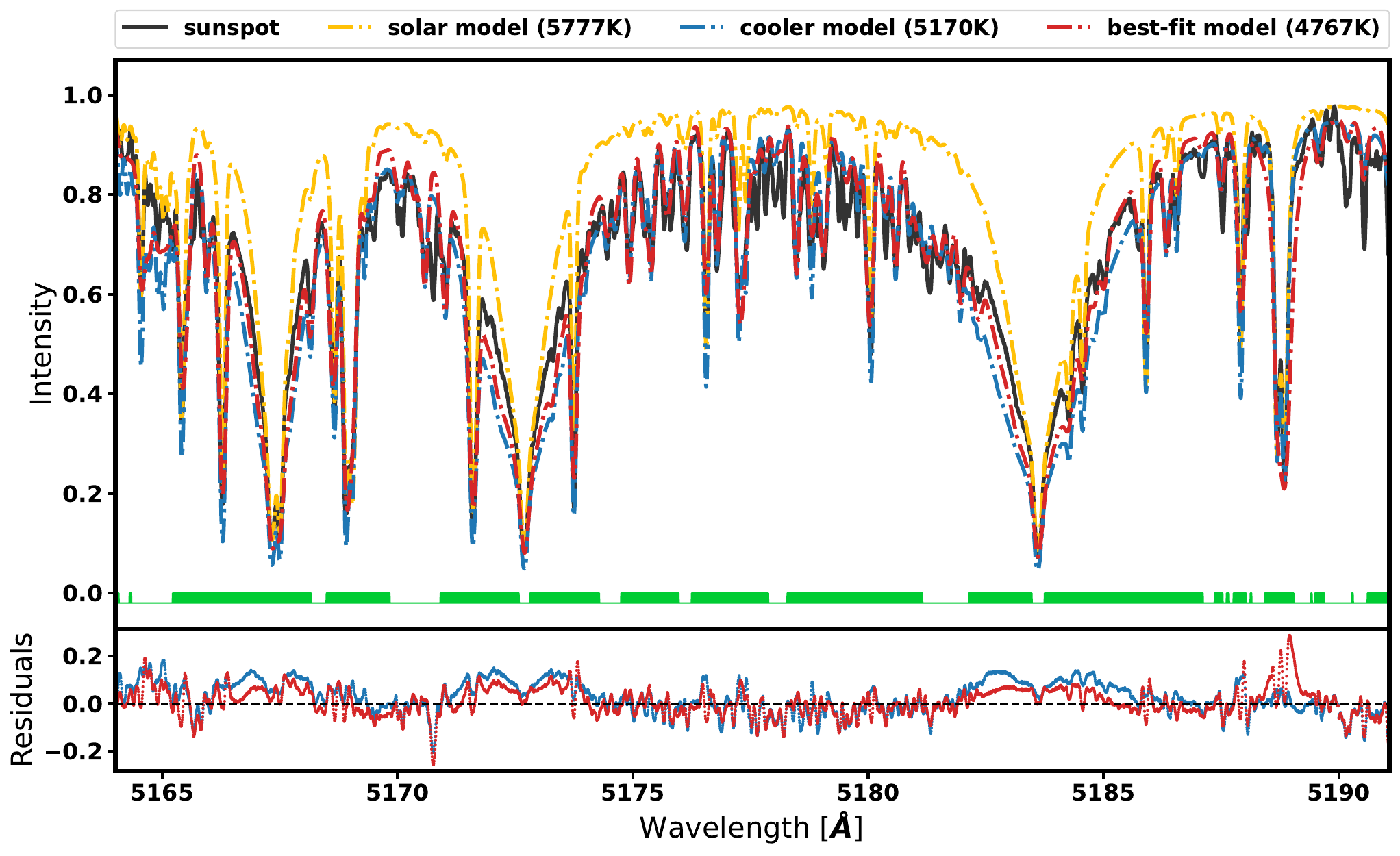}
\caption{\label{fig:general} \emph{Top plot}: A comparison between the observed Sunspot spectrum (\textit{black}) with three different models synthesized by SME. The three models are Solar (\textit{yellow}), cooler (\textit{blue}), and best-fit (\textit{red}). The green bars at the bottom of the plot indicate regions used to determine the goodness of fit between the model and the observed spectrum. \emph{Bottom plot}: Residuals between the observation and cooler model (\textit{blue}) with residuals between the observation and best-fit model (\textit{red}).} 
\end{figure*}

\section{Spectral Fitting Method} \label{sec:sme}

We used Spectroscopy Made Easy (SME, \citep{Valenti_Piskunov_1996}) to fit spectral models to the observed Sunspot spectrum. SME incorporates a stellar atmosphere grid, a line list, and a radiative transfer code to create a model spectrum based on physical parameters which can be fixed or free in any combination. These physical parameters include effective temperature (\teff), surface gravity (\logg), metallicity ([M/H]), rotational velocity (\vsini), macroturbulence (\vmac), microturbulence (\vmic), radial velocity (\vrad), and a continuum scaling factor (Cscale). The continuum scaling factor fits a line to the continuum of the observed spectrum to account for any residual trends.

The line list used in our work, which includes both atomic and molecular line data, was taken from \citet{Brewer_2016} and includes $> 7000$ lines from 5162{\AA} to 7800{\AA} in 20 wavelength segments totalling 350{\AA}. We were limited by the spectral range of the observation at the red end to 6664 {\AA}. This allowed us to model 268{\AA} of the observation in 14 wavelength segments. The lines were chosen to return accurate temperatures and gravities for a selection of stars that had asteroseismic observations; some regions were chosen to better constrain the abundances of 15 elements. Most of the temperature and gravity information is in the blue end of the line list.  Losing the red end will therefore only primarily affect the accuracy of the abundances for some elements \citep{2015ApJ...805..126B}.

The \citet{Brewer_2016} line list was originally taken from the Vienna Atomic Line Database 3 \citep[VALD-3;][]{Ryabchikova_2015} and calibrated against the quiet Solar spectrum \citep{Wallace_2011}. Approximately 20\% of the lines were discrepant from the quiet Solar spectrum in line-center, line-depth, or line breadth, requiring changes to wavelength, oscillator strength, or Van der Waals broadening coefficient respectively. These parameters were fit for while holding the other stellar parameters fixed at the standard Solar values (refer to Solar model in Table \ref{table:1}). Features in the Solar spectrum that had no corresponding line data or that were poorly fit were masked out of the fitting regions. More details regarding this calibration method can be found in Section 3.2 of \cite{Brewer_2016}.

As in \citet{Brewer_2016}, we used the Kurucz 2012 model atmospheres \citep{Castelli_2003} for all of our spectral models. To reduce the effects of telluric contamination, we masked out all spectral regions where the telluric atlas \citep{Wallace_2011} was less than 99\% of the continuum.

We made three spectral models using SME; the stellar parameters for each model are given in Table \ref{table:1}. Reported uncertainties are empirically determined uncertainties from \citet{Brewer_2016} as derived from integrated light spectra of multiple observations of the same stars. For the two models where we fit for parameters, we had to include uncertainties for the observed umbral spectrum.  Given the typical S/N of observations from the McMath-Pierce Solar Telescope, we can assume that the uncertaines for the Sunspot spectra would be no worse than the Keck HIRES spectrum we later use for empirical comparison, and thus set them to $0.004$ divided by the normalized intensity.

Our Solar model was constructed with all parameters fixed at Solar values. This was designed to create a model of the quiet Solar surface to compare with the Sunspot observation. By construction, it is a good fit to the NSO quiet Solar atlas \citep{Brewer_2016,Wallace_2011}.

The cooler model was fit leaving only temperature as a free parameter, while all other parameters were fixed at Solar values. This model was constructed to compare a spectrum of a cooler Sun to the Sunspot observation and to compare with average Sunspot temperatures from literature. 

Our best-fit model was made following the iterative fitting procedure of \citet{Brewer_2016} due to its ability to derive precise temperatures, abundances for 15 elements, and surface gravities consistent with those from asteroseismology \citep{2015ApJ...805..126B}. The additional free parameters allows our model to attempt to fit line shapes in the Sunspot spectrum that are not well described by just a cooler temperature.
In step 1 of the best-fit modeling procedure, the global parameters \teff, \logg, \vmac, [M/H], and the individual abundances of Ca, Si, and Ti are set as free parameters, with all other abundances scaled to [M/H]. 
We set the initial parameters to Solar except for \teff\ $= 5500$ and used SME to fit a model.  
We perturbed the resulting model by $\pm 100$K, and re-fit. In step 2, we fixed those global parameters at the weighted means from step 1, and allowed the abundances of 15 elements (C, N, O, Na, Mg, Al, Si, Ca, Ti, V, Cr, Mn, Fe, Ni and Y) to vary.
We then repeated steps 1 and 2 but now with this new abundance pattern and the initial parameters set to the values from step 2.  Finally, we fixed all parameters except \vsini, used the \vmac\ relation from \citet{Brewer_2016} to fix the \vmac, and solved for \vsini. We discarded this step since the Sunspot is localized on the Solar surface and should incur no significant rotational broadening. Final parameters are collected in Table \ref{table:1} as our 'best-fit' model.

\section{Results and Analysis} \label{sec:RandA}

\begin{figure}[htb!]
\includegraphics[width=\linewidth]{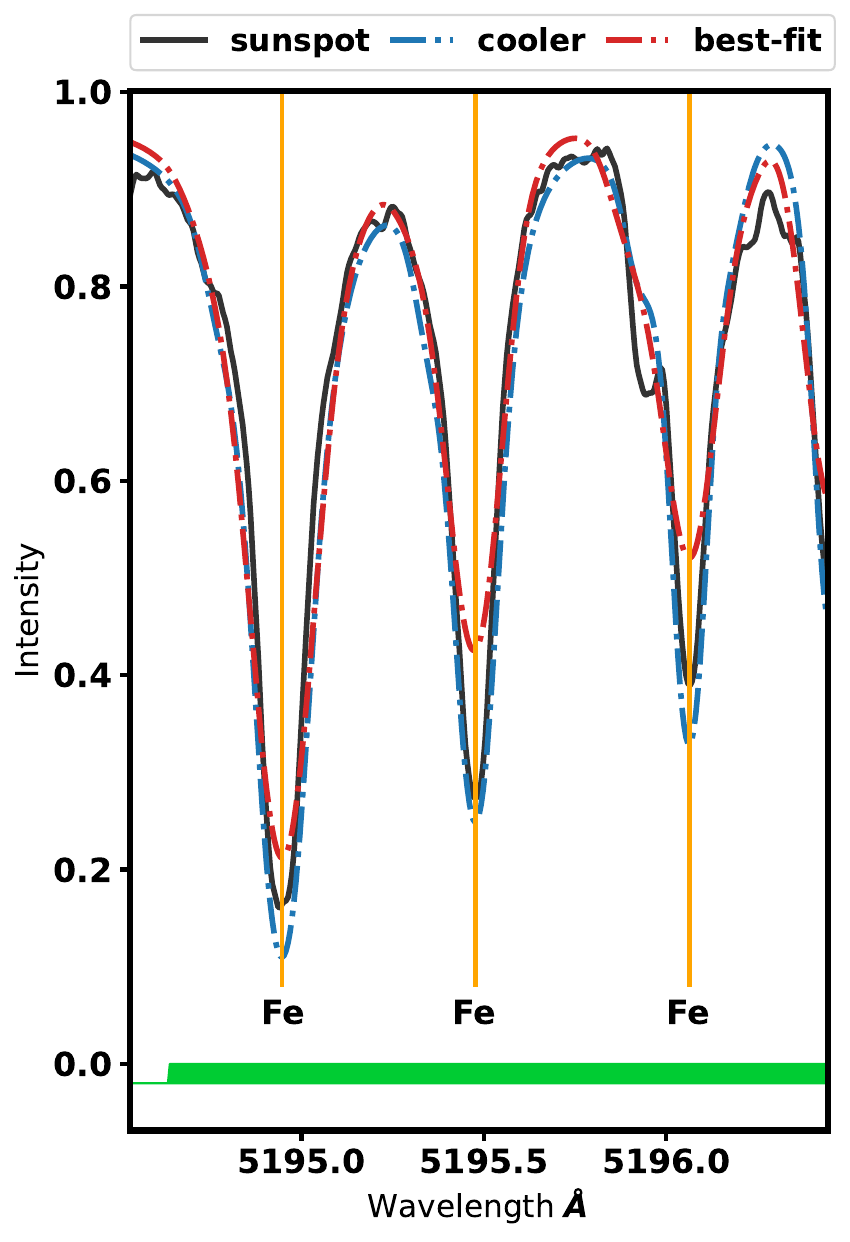}
\caption{\label{fig:goodeg} An example in which the cooler (\textit{blue}) and best-fit (\textit{red}) models fit the observation (\textit{black}) well. The spectral lines seen in this wavelength range are all iron lines, with their expected central wavelengths marked by vertical, orange lines. All three lines have simple structures; they are visually symmetrical and do not exhibit significant blending or splitting. Both models fit the first iron line fairly well in both core depth and wing shape. The cooler model does slightly better than the best-fit model at reproducing the depth of the other two lines.}
\end{figure}

\begin{figure*}[htb!]
\includegraphics[width=\linewidth]{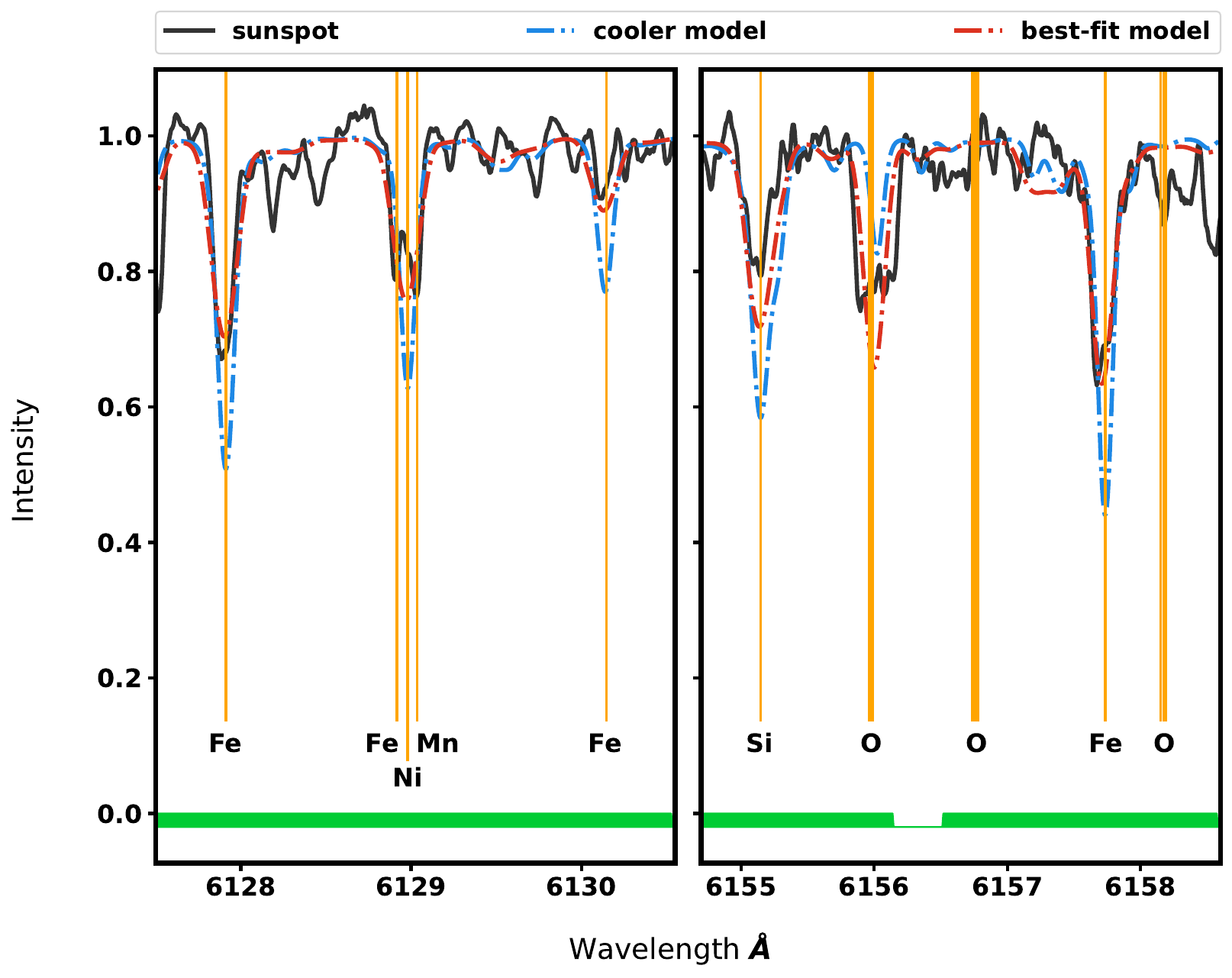}
\caption{\label{fig:badeg} Two examples of where the cooler and best-fit models fit the observation poorly. The vertical orange lines indicate the most prominent spectral lines within each wavelength range. In both plots, the cooler model overestimates the core depths of most lines, while the best-fit model does slightly better at reproducing them. The observation shows crowds of spectral lines in the areas for which the models assume a smooth continuum. On the \textit{left} plot at 6129{\AA}, where the models depict a single, strong nickel line, the observation exhibits iron and manganese lines adjacent to the nickel line. The \textit{right} plot includes a silicon line and the oxygen triplets. Within the Sunspot observation, the first oxygen line at 6155{\AA} is imbeded in a well-like structure, which both models fail to accurately reproduce the depth and width of.}
\end{figure*}

\subsection{SME Spectral Synthesis} \label{sec:result_SME}

Spectral synthesis performed under the assumption of LTE for three different models (as described in Section \ref{sec:sme}) yielded the results seen in Figure \ref{fig:general}, which shows one spectral segment of the 14 total segments of data. The green bar at the base of the top plot represents the mask of well-calibrated lines \citep{Brewer_2016}; these are the regions used to calculate $\chi^2$ to determine method performance during the fitting. The bottom plot shows the residuals between the cooler and best-fit models and the observation. 

The Solar model (\textit{yellow}) and the observation (\textit{black}) have significant differences in the depth and width of the lines throughout the segment (Figure \ref{fig:general}). The Sunspot observation has much deeper lines with wider wings. Additionally, the observation shows a greater number of lines in areas where the Solar model has a smooth continuum.

The cooler model, made by solving for temperature as a free parameter, yielded a global effective temperature of \teff\ = 5170K. This is a reasonable value considering that a Sunspot is expected to have a temperature 500-1500K cooler than the quiet Solar surface \citep{Solanki_2003}. Compared to the Solar model, the cooler model shows broad agreement with the observation as can be seen in Figure \ref{fig:general}, where the intensities, width, and cores of lines in the cooler model, shown in \textit{blue}, are now much closer to the observation. 

The best-fit model resulted in changes to a majority of the stellar parameters, as shown in Table \ref{table:1}, row 3. This model has a significantly cooler effective temperature of 4767K, which again lies within the expected temperature range of a Sunspot. The model also has lower surface gravity and metallicity than Solar values.

Broadening from \vmac\ and \vsini\ are largely degenerate.  Prior to the final step of our iterative fitting procedure, the broadening from \vmac\ and \vsini\ are folded into a single `rotational broadening' parameter using the macroturbulence broadening kernel.  Since the Sunspot is localized on the Solar surface, there should not be much broadening due to \vsini.  Although we found this broadening to be several km/s higher than it is for the quiet Sun (Table \ref{table:1}), some of this could be due to the varying resolution across the observed spectrum.

In comparison to the Solar model, the best-fit model agrees significantly better with the observation, as seen in Figure \ref{fig:general}. When compared to the cooler model, the best-fit model does slightly better at fitting the core depths and wings of lines (Figure \ref{fig:general}). We used the Akaike Information Criterion \citep[AIC;][]{Akaike.1974} to compare the fits of the two models.  AIC favors models with better fits, but penalizes overly complex models with more free parameters.  Our best-fit model is strongly favored over the cooler model with an AIC 20 points lower.  However, visual inspection shows many regions in both models that do not accurately reproduce the observed line shapes.

Some regions are well-fit in both the cooler and best-fit models (Figure \ref{fig:goodeg}). The spectral lines seen in this plot have very simple structure and do not present significant line shape changes such as blending and splitting. While both models fit quite well in this wavelength range, the cooler model does slightly better at reproducing the core depths of the lines while the best-fit model does slightly better at fitting the wings of the lines. Despite these well-fit areas, there are some large deviations that can be seen in the residuals between the best-fit and observation (Figure \ref{fig:general}, bottom).

Although the cooler model of 5170K and the best-fit model of 4767K are in better agreement with the observed Sunspot spectrum, there are many lines that do not adequately fit. The differences between the models and the observation are shown in Figure~\ref{fig:general} as a residual plot at the \textit{bottom}, where the positive values indicate overestimated line depth and the negative values indicate underestimated line depth.

Additional deviations can be seen in Figure \ref{fig:badeg}. Despite both panels showing wavelength ranges where we are confident in the line parameters, the cooler and best-fit models are unable to accurately reproduce the complex line features seen in the observation. In the \textit{left} panel of Figure \ref{fig:badeg}, the cooler model over estimates the line depths of the three most prominent (iron and nickel) lines. While the best-fit model is able to accurately reproduce the line depth for the first and third most prominent line, there is added complexity in the structure of the second line. In the observed spectrum, the second line shows splitting at the core and broadened wings. Here, both models have one prominent nickel line.  However, the observation indicates adjacent, strong iron and manganese lines, which seem to have blended with the nickel line. The best-fit model tries to capture this structure in its entirety by broadened wings and shallower core. 

For the \textit{right} plot of Figure~\ref{fig:badeg}, both models fail to accurately reproduce the line shapes seen in the observation in both depth and width. The models \textit{overestimate} the depth and width of the silicon line at 6155 {\AA}. At 6156 {\AA}, the observation shows a combination of complex features, making this particular region have a well-like structure. The cooler model fails to capture this region in shape, width, and depth, while the best-fit model, again, attempts to fit the region with one broadened line.

\subsection{Magnetic Field Strength} \label{sec:result_gauss}

\begin{figure}[htb!]
\includegraphics[width=\linewidth]{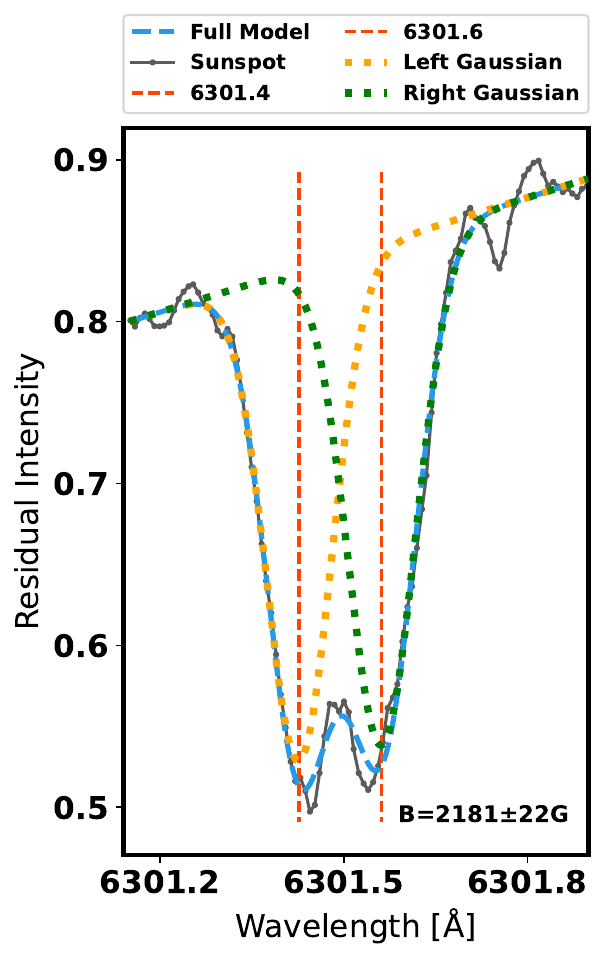}
\caption{\label{fig:zeeman} Zeeman splitting of an iron line with a central wavelength of 6301.5{\AA}. The two dashed, red, vertical lines indicate the symmetrical splitting and were found to be $\pm$0.067{\AA} from the central wavelength. The splitting was measured using a double Gaussian fit for which the full model is shown with the blue line. Using equation \ref{eq:2}, it calculated that a magnetic field strength of roughly 2181G is necessary to cause a splitting of this degree.}
\end{figure}

Magnetically sensitive lines (both atomic and molecular) can undergo line shapes changes due to Zeeman splitting, an NLTE effect. This effect describes the splitting of an absorption line into multiple absorption features in the presence of an intense magnetic field. This arises due to each energy level, or state of an atom, having a magnetic moment.  When an atom is in a magnetic field, each single energy state is split into closely spaced states that are shifted in wavelength \citep{Landstreet_2015}. This wavelength shift can be expressed as

\begin{equation}\label{eq:1}
    \Delta\lambda_{ij} = 4.67 \cdot 10^{-13} \lambda_0^2 (g_L m_{Ji} - g_L m_{Jj}) B 
\end{equation}

where \emph{g}\textsubscript{L} is the dimensionless Landé factor, $\lambda_{0}$ is the rest-frame wavelength in units of {\AA}, \emph{B} is the magnetic field strength in Gauss, and (\emph{m}\textsubscript{Ji} \textminus \emph{m}\textsubscript{Jj}) represents allowed transitions in the magnetic moment and is equal to $\pm$1. This expression (eq. \ref{eq:1}) can be re-arranged into the following,

\begin{equation}\label{eq:2}
    B = \frac{\Delta\lambda_{ij}}{4.67 \cdot 10^{-13} \lambda_0^2 g_L} 
\end{equation}

to solve for the magnetic field strength based on the wavelength shift of Zeeman splitting.

We found numerous potential splitting features throughout the Sunspot observation. To evaluate the magnetic field strength, we focus on an iron line at 6301{\AA}, shown in Figure \ref{fig:zeeman}, for two reasons. One, it has relatively simple structure and is a visibly clean splitting. Two, there are no atomic or molecular lines close to the split structure, and therefore it is unlikely that the structure has been impacted by blending with broadened adjacent lines. However, the latter must be taken with caution, as there could potentially be molecular lines around the structure that are unaccounted for in our line list. 

The central wavelength ($\lambda\textsubscript{0}$) for the iron line is 6301.5{\AA} and the splitting was measured to be $\Delta\lambda$ = 0.067 $\pm$ 0.001 {\AA}. To measure the separation, we fit a simple two-Gaussian model with a slope to account for the sloping continuum in the region. The slope, offset, central wavelength, and the centers, amplitudes, and widths of the Gaussians were all free parameters.  The Gaussian centers were constrained to be equadistant from the central wavelength. The best fit Gaussian model is shown in \textit{blue} in Figure \ref{fig:zeeman}. The Landé factor is \emph{g}\textsubscript{L} = 1.67 taken from VALD-3 \citep{Ryabchikova_2015}. The magnetic field strength necessary to cause this degree of splitting is found to be 2181 $\pm$ 22G. This is within the range of the average measured magnetic field strength of 1500 to 3500G \citep{Livingston_2015, Pevtsov_2021}. We performed the same analysis on a Zeeman split titanium line at 6017.5{\AA}. This line is much shallower, about 4\% of the continuum, but similarly isolated. We determined a field strength of 2207 $\pm$ 32G, which is consistent with the field strength returned for the deeper Fe line.

\section{Discussion} \label{sec:dis}

\begin{figure*}[htb!]
\includegraphics[width=\linewidth]{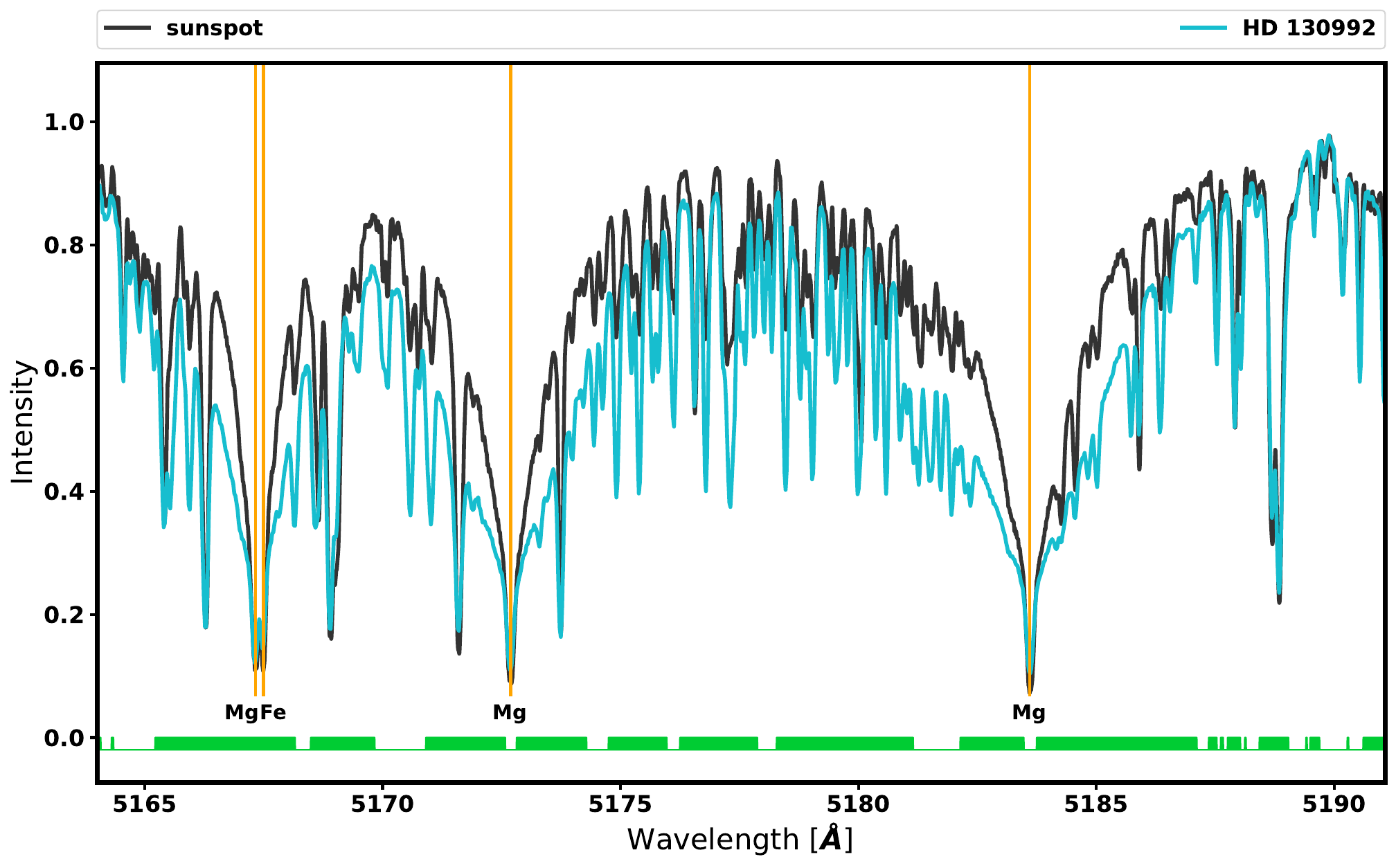}
\caption{\label{fig:ss_cool1} A comparison between the Sunspot observation (\textit{black}) and the cool star (HD 130992) observation (\textit{light blue}). The vertical, orange lines indicate the four most prominent spectral lines found in this wavelength range. The reduced pressure broadening in the strong lines of the Sunspot are evident in the magnesium triplet lines. (Continuum offset (+0.06) was added to the cool star.)}
\end{figure*}

\begin{figure}[htb!]
\includegraphics[width=\linewidth]{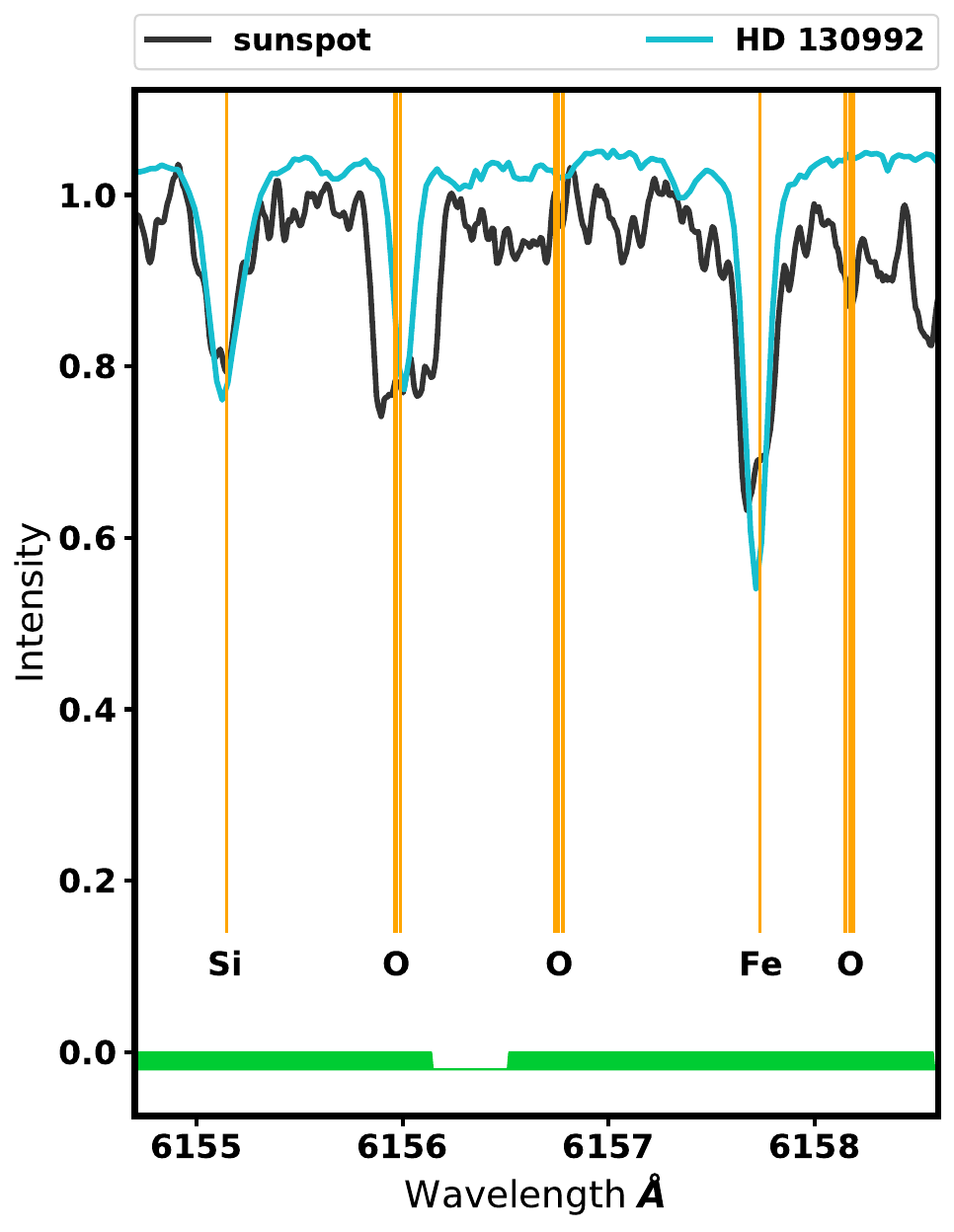}
\caption{\label{fig:ss_cool_O} A closer look at the differences between the Sunspot spectrum and a cool star (HD 130992) spectrum for wavelength region 6155-6158{\AA}. The peculiar change to the oxygen line at 6156{\AA} in the Sunspot spectrum cannot be found in HD 130992. Weaker lines are much more strengthened in the Sunspot than in HD 130992. (Continuum offset (+0.06) was added to the cool star.)}
\end{figure}

We have investigated the ability of a cooler Sun and best-fit LTE models to reproduce the Sunspot spectrum using SME. Our spectral analysis has shown the complicated nature of Sunspots, and the inability of a simple 1D-LTE model to accurately model the spectral line features seen in observation. Furthermore, comparison between the Sunspot observation against a cool star observation indicate that complex spectral line distortions found in the Sunspot spectrum are likely induced by its magnetic field.

\subsection{Changes in Photospheric Conditions in Sunspots} \label{sec:dis_model_limit}

A detailed comparison between the three LTE models against the Sunspot observation indicates the significant differences in physical conditions and stellar parameters between a quiet Sun and Sunspot. Our three models made using SME are the Solar model, cooler model, and best-fit model. The Solar model had all parameters set at Solar values, and was used to compare the observation against a quiet Sun. The cooler model is a model with a cooler effective temperature but is otherwise identical to the quiet Sun. The best-fit model was made to extract stellar parameters of the Sunspot and allowed for all parameters to be varied. Both the cooler and best-fit models significantly decreased in \teff\ in comparison to the Solar effective temperature of 5777K. Nearly all of the parameters in the best-fit model experienced significant changes from Solar values (see Table \ref{table:1}). 

\subsubsection{Temperature}
In comparison to the Solar model and the quiet Sun, which has a surface temperature of 5777K, the cooler and best-fit models yielded a much cooler global fit temperature of 5170K and 4767K, respectively, for the Sunspot. Both temperatures are within the range of what can be expected for a Sunspot temperature \citep{Solanki_2003}. The Solar model clearly does not fit well with the Sunspot observation (Figure \ref{fig:general}). The difference in depth is the major deviation between the Solar model and the Sunspot observation, seen across the spectrum (Figure \ref{fig:general}), as the cooler temperature of the Sunspot causes most neutral lines to increase in depth. With the cooler and best-fit model accounting for the cooler surface temperature expected for a Sunspot, the two models fit the line depths of the observation significantly better than the Solar model.

\subsubsection{Gravity}
A lower \teff\ is able to capture most of the line depth changes, but the fixed \logg\ of the cooler model now overestimates the broadening of the line wings.  The strong magnetic field of a Sunspot supresses convection, lowering the pressure broadening. This results in line profiles more like those of a subgiant, with narrow line wings indicative of less pressure broadening due to lower surface gravity. As SME cannot incorporate magnetic field effects, the best-fit model attempts to account for the change in gas pressure by reducing the surface gravity in the stellar parameters. The residuals of the best-fit model tend to be closer to zero than those of the cooler model in the wings of strong lines (Figure \ref{fig:general}, bottom).

\subsubsection{Molecular Lines}
Cooler stellar surface temperatures allow for increased abundances of some molecular species \citep{Gaur_1973, Pande_1975, Asplund_2021} resulting in their lines becoming enhanced in Sunspots \citep{Moore_1966}. Our modeling simultaneously fits molecular and atomic lines; therefore, atoms forming into molecules should not have a large affect on our measured elemental abundances. For instance, the Solar photospheric spectrum contains some shallow molecular CO and C$_2$ lines. Because the Sun is oxygen rich, we expect a decreasing temperature to increase the abudance of CO, but decrease the abundance of C$_2$.  This results in slightly shallower lines of molecular carbon, while the CO lines get stronger.  At the same time, many neutral carbon and oxygen lines will be strengthened as excited atoms shift to lower energies.  Because we include lines of C, O, CO and C$_2$ in our model, these should result in no net change in abundance.

However, notice that for both the cooler and best-fit models, they often assume a continuum in areas where there are clearly spectral lines in the observation (Figure \ref{fig:badeg}). Most of these missing spectral lines are likely molecular lines, many of which are unaccounted for in our line list. Although our line list does include some molecular lines, those may also not be accurately tuned. Because the Sun is used as our calibrator, the temperature is too high for most molecules, and thus the line parameters are difficult to verify. Completeness and accuracy have been a major challenge in developing molecular line lists for astrophysical modeling, although there is significant work being done to create a molecular line list for, much cooler, exoplanet atmospheres \citep[e.g.][]{Rothman_2009, Tennyson_2024}. Additional observations to obtain spectra of resolved Sunspots could be a good source of information on molecular lines strengths and broadening under astrophysical conditions. 

\subsection{Magnetic Field Effects on Line Shape}
In addition to the cooler temperature, lower gravity, and increased abundance of molecular lines, another important factor that can cause significant discrepancies between the models and the Sunspot observation are the NLTE effects induced by the magnetic field of the Sunspot. The magnetic field can cause line distortions due to the Zeeman effect. The Zeeman effect can cause strengthening of saturated lines, otherwise known as the Zeeman or magnetic intensification \citep{Babcock_1949, Stift_2003}. It can also cause lines to be broadened (a consequence of when the width of the magnetic splitting is less than the unmagnetized line width) or split (if the magnetic field is strong enough), where the latter is often called Zeeman splitting. We discuss Zeeman splitting further in Section~\ref{sec:dis_zeeman} below.

Changes in line shape in the form of broadening, splitting, and intensification due to the Zeeman effect can be seen throughout the spectrum. Several prominent lines show splitting at their cores (e.g. the Si and Fe lines in the right panel of Figure \ref{fig:badeg}). The more complex the structure of the lines, the more likely it is that a combination of different factors are causing the change in line shapes, which clearly a simple LTE model is not able to account for. 

As is shown in Figure \ref{fig:badeg}, both cooler and best-fit models fail at reproducing the spectral lines that have undergone complex changes in line shape. Within Figure \ref{fig:badeg}, the cooler model often overestimates the line depth for the most prominent lines. This indicates the existence of a different NLTE effect causing the unintuitive reduction in the core depth. In regards to the best-fit model, it does well at fitting the depths of the most prominent lines seen in Figure \ref{fig:badeg}. However, for lines featuring broadening, splitting, and/or blending with weaker lines, the model tries to adapt to these changes by broadening a single line to fit the width of the feature. This adaptation has led to significant changes in the stellar parameters of the best-fit model. 

We also saw evidence of spectral differences that cannot easily be explained by a single parameter change. The especially large distortion in the observation seen at 6156{\AA} in Figure \ref{fig:badeg}, where the region in and around an oxygen line has changed into a well-like structure, is a great example of a line undergoing multiple different broadening, splitting, blending, and magnetic effects. Both the cooler and best-fit models fail to reproduce this structure. Furthermore, the magnetic field affects lines of the same element differently depending on their wavelength. Despite the oxygen triplet (6156-6158{\AA} of Figure \ref{fig:badeg}) being in close proximity with each other, the line shapes are drastically different with 6156{\AA} undergoing the strongest change.

\subsubsection{Velocity Shift by Zeeman Splitting} \label{sec:dis_zeeman}

The splitting of a magnetically sensitive spectral line into multiple absorption features in the presence of an intense magnetic field is called Zeeman splitting. There is direct correlation between the magnetic field strength of a Sunspot with the magnitude of split of an absorption line, as seen in equation \ref{eq:1} in Section \ref{sec:result_gauss}. We can thus use our observation to evaluate the field strength of this particular Sunspot.

In Section \ref{sec:result_gauss} we evaluate the magnetic field strength, $B$, using Zeeman split titanium and iron lines at 6017.5{\AA} and 6301.5{\AA} respectively, finding $B$ to be $\sim 2200$G. This is within the range of the average measured Sunspot magnetic field strengths of 1500 to 3500G \citep{Livingston_2015, Pevtsov_2021}. 

Zeeman splitting induced by the magnetic field of a Sunspot can also introduce spurious RV shift. The equation for the wavelength shift due to Zeeman splitting, shown in section \ref{sec:result_gauss}, can be rearranged to describe the degree of velocity shift due to Zeeman splitting

\begin{equation}\label{eq:3}
    \Delta\textit{v} = 1.4 \cdot 10^{-7} \lambda_0 g_L B
\end{equation}

where $\Delta$\textit{v} has units of km/s. Equation \ref{eq:3} shows that longer wavelengths experience stronger effects of Zeeman splitting. Within the visible range of 5164-6664{\AA}, a Sunspot with magnetic field strength of 2174G, as calculated for the iron line in the section \ref{sec:result_gauss}, can cause a veloocity shift of roughly 2.5-3.5 km/s. The translation from $\Delta\lambda$ to $\Delta\textit{v}$ is also not as straightforward as it seems, as Zeeman splitting can affect different elements to varying degrees; while some lines may have clearly separated splitting, others may only appear broadened. We cannot expect an overall uniform velocity shift throughout a spectrum. Additionally, only a fraction of the integrated disk spectrum will have these line shifts. However, even subtle changes in line shape would affect our ability to precisely measure planet induced radial velocities. Further details on Zeeman splitting can be found in literature including \cite{Ruslan_2007, Reiners_2012, Landstreet_2015}.

\subsection{Empirical Comparison} \label{sec:empirical}

As noted above, our 1D-LTE models are missing some details that may hinder our analysis. Our line-list, especially for molecular lines, may be incomplete or inaccurate. Even small errors in line parameters can be magnified at temperatures far from Solar. Additionally, NLTE effects could be causing some of the differences we note. We can overcome all of these limitations by instead comparing the Sunspot to the observed spectrum of a cooler star. One complication of comparison to a disk integrated spectrum is rotational broadening that won't be present in the Sunspot; to minimize this we can choose a slowly rotating star.

We use the star HD 130992 as an empirical comparison. HD 130992 was chosen because it has the same temperature as the effective temperature yielded by our best-fit model of the Sunspot and nearly Solar metallicity and surface gravity. Additionally, the total broadening from \vmac\ and \vsini\ is only 1.7 km/s, with \vsini=0.7$\pm 0.5$ km/s.  This slow rotation will add only minimal broadening to the lines. The observed spectroscopic comparison between the Sunspot and HD 130992 are shown in Figures \ref{fig:ss_cool1} and \ref{fig:ss_cool_O}. 

While there are many similarities between the two observations, the absorption features in HD 130992 are generally stronger than they are for the Sunspot (Figure \ref{fig:ss_cool1}). This is due to prominent lines experiencing stronger effects of pressure broadening for HD 130992, which has a slightly higher surface gravity (\logg\ $= 4.51$) than the Sun. This causes the broadened wings to blend with neighboring lines. Though the Sunspot and HD 130992 have similar effective temperatures, the best-fit model of the Sunspot yielded a significantly lower surface gravity (\logg=3.68 vs. 4.51 for HD 130922), helping it better fit the narrower lines of the Sunspot observation. 

Another significant difference can be seen in Figure \ref{fig:ss_cool_O}, which covers the same wavelength range as the \textit{right} plot of Figure \ref{fig:badeg}. The peculiar change of the oxygen line at 6156{\AA} into a well-like structure in the Sunspot observation is not seen in HD 130992. This demonstrates that we are likely not lacking lines in our line list, but that the change is due entirely to the magnetic splitting of the line. Furthermore, many of the weak lines in the Sunspot spectrum are significantly strengthened in comparison to HD 130992. These discrepancies are also likely caused by the intense magnetic fields in the Sunspot. HD 130992 has very similar stellar parameters to the Sun with the exception of temperature, and has a low R'\textsubscript{HK} value indicating low activity. The magnetic field in the Sunspot strengthens the normally weak lines while also causing severe distortions in some of the lines via a combination of broadening and splitting. These phenomena are not observed in the spectrum of the cooler star HD 130992.

\section{Conclusion} \label{sec:con}

We have studied whether a cool Sun model is able to accurately model a Sunspot observation. We compared the full Sunspot spectrum in the visible to three LTE models made using SME. The Solar model, made with all model parameters fixed at Solar values, showed how different the Sunspot observation is in comparison to the quiet Solar surface. The model showed visible differences in line depth against the observation throughout the spectrum. The cooler model solved for temperature as a free parameter, leaving all other parameters fixed at Solar values. This model was essentially a cool Sun model and yielded a global fit effective temperature of 5170K. The best-fit model was made through an iterative fitting process and yielded a globl fit effective temperature of 4767K as well as a much lower surface gravity and metallicity. The temperatures of both cooler and best-fit models are within the expected temperature range for an average Sunspot. 

There exists lines with simple structure (i.e. symmetrical with negligible blending, broadening, and splitting) for which changing the temperature of the models adequately accounted for the different core depths and wings of the lines in the Sunspot observation. However, we also observed many lines with complex structures (i.e. not symmetrical and undergoing strengthening, blending, broadening and/or splitting) that likely arose due to a combination of different NLTE effects of the Sunspot. One significant NLTE effect is the Zeeman effect which is induced by the Sunspot's magnetic field. The Zeeman effect can cause significant distortions in line shapes via strengthening, broadening, and splitting. Our models were unable to reproduced these distortions due to the inability of SME to incorporate magnetic effects. Additionally, the Sunspot observation shows indications of increased occurence of molecular lines which our models struggle to fit. This could be due to our line list being incomplete or not accurately tuned. 

Due to finding a myriad of line shape variations across the observed Sunspot spectrum, we were motivated to conduct an empirical comparison of the Sunspot observation against a spectrum of a cool star, which does not suffer from lack of molecular lines nor different NLTE effects apart from magnetic ones. HD 130992 was chosen due its similarities in stellar parameters with that of the Sun and because it has the same effective temperature as that yielded by our best-fit model of the Sunspot observation. Spectroscopic comparison of the Sunspot and HD 130992 show significant discrepancies in line shape. The spectral lines in the Sunspot observation experience weaker line broadening than for HD 130992, likely from the suppression of convection yielding lower pressure broadening in the Sunspot. Spectral lines with complex structure found in the Sunspot spectrum were not seen in HD 130992, strengthening the theory that these were induced by the magnetic field of the Sunspot and not by missing lines in our models. 

We calculated the magnetic field strength for the Sunspot at $\sim 2200$G using the Zeeman splitting of an iron line and a titanium line. This is within the range of what can be expected for an average Sunspot. A Sunspot of this field strength is enough to cause a velocity shift of roughly 2.5-3.5 km/s in the visible. The impact of Sunspot activity on RVs and the failure of a simply cooler Sun model to replicate the Sunspot spectrum emphasizes the necessity for new stellar activity modeling and Sunspot activity mitigation methods for RV data, especially when trying to detect small planets using EPRV instruments. Additional observations of Sunspots of varying sizes and magnetic field strengths will enable us to constrain the variability of other stars and help refine our understanding of molecular line strengths and NLTE effects induced by Sunspot activity.

\begin{acknowledgments}
C.K. \& J.M.B. gratefully acknowledge the support for this research from NASA Grant 80NSSC21K0009, NASA XRP 80NSSC21K0571 and NSF AAG award 2307467. Additionally, C.K. is grateful to R.M.R., J.L., and E.C.G. for helpful discussions which improved the paper. The Flatiron Institute is a division of the Simons Foundation. Support for this work was provided by NASA through the NASA Hubble Fellowship grant HST-HF2-51569 awarded by the Space Telescope Science Institute, which is operated by the Association of Universities for Research in Astronomy, Incorporated, under NASA contract NAS5-26555. This work has made use of the VALD database, operated at Uppsala University, the Institute of Astronomy RAS in Moscow, and the University of Vienna.
\end{acknowledgments}

\bibliography{Sunspot_paper}

\begin{thebibliography}{}
\expandafter\ifx\csname natexlab\endcsname\relax\def\natexlab#1{#1}\fi
\providecommand{\url}[1]{\href{#1}{#1}}
\providecommand{\dodoi}[1]{doi:~\href{http://doi.org/#1}{\nolinkurl{#1}}}
\providecommand{\doeprint}[1]{\href{http://ascl.net/#1}{\nolinkurl{http://ascl.net/#1}}}
\providecommand{\doarXiv}[1]{\href{https://arxiv.org/abs/#1}{\nolinkurl{https://arxiv.org/abs/#1}}}

\bibitem[{Akaike(1974)}]{Akaike.1974}
Akaike, H. 1974, IEEE Transactions on Automatic Control, 19, 716, \dodoi{10.1109/tac.1974.1100705}

\bibitem[{{Asplund} {et~al.}(2021){Asplund}, {Amarsi}, \& {Grevesse}}]{Asplund_2021}
{Asplund}, M., {Amarsi}, A.~M., \& {Grevesse}, N. 2021, \aap, 653, A141, \dodoi{10.1051/0004-6361/202140445}

\bibitem[{{Babcock}(1949)}]{Babcock_1949}
{Babcock}, H.~W. 1949, \apj, 110, 126, \dodoi{10.1086/145192}

\bibitem[{{Balthasar}(1999)}]{Balthasar_1999}
{Balthasar}, H. 1999, \solphys, 187, 389, \dodoi{10.1023/A:1005131927915}

\bibitem[{{Basant} {et~al.}(2025){Basant}, {Luque}, {Bean}, {Seifahrt}, {Brady}, {Zhao}, {Brown}, {Das}, {St{\"u}rmer}, {Kasper}, {Gupta}, \& {Stef{\'a}nsson}}]{Basant_2025}
{Basant}, R., {Luque}, R., {Bean}, J.~L., {et~al.} 2025, \apjl, 982, L1, \dodoi{10.3847/2041-8213/adb8d5}

\bibitem[{{Boisse} {et~al.}(2011){Boisse}, {Bouchy}, {H{\'e}brard}, {Bonfils}, {Santos}, \& {Vauclair}}]{Boisse_2011}
{Boisse}, I., {Bouchy}, F., {H{\'e}brard}, G., {et~al.} 2011, \aap, 528, A4, \dodoi{10.1051/0004-6361/201014354}

\bibitem[{Brewer {et~al.}(2015)Brewer, Fischer, Basu, Valenti, \& Piskunov}]{2015ApJ...805..126B}
Brewer, J.~M., Fischer, D.~A., Basu, S., Valenti, J.~A., \& Piskunov, N. 2015, The Astrophysical Journal, 805, 126, \dodoi{10.1088/0004-637x/805/2/126}

\bibitem[{{Brewer} {et~al.}(2016){Brewer}, {Fischer}, {Valenti}, \& {Piskunov}}]{Brewer_2016}
{Brewer}, J.~M., {Fischer}, D.~A., {Valenti}, J.~A., \& {Piskunov}, N. 2016, \apjs, 225, 32, \dodoi{10.3847/0067-0049/225/2/32}

\bibitem[{{Cao} \& {Pinsonneault}(2022)}]{Cao_2022}
{Cao}, L., \& {Pinsonneault}, M.~H. 2022, \mnras, 517, 2165, \dodoi{10.1093/mnras/stac2706}

\bibitem[{{Castelli} \& {Kurucz}(2003)}]{Castelli_2003}
{Castelli}, F., \& {Kurucz}, R.~L. 2003, in Modelling of Stellar Atmospheres, ed. N.~{Piskunov}, W.~W. {Weiss}, \& D.~F. {Gray}, Vol. 210, A20, \dodoi{10.48550/arXiv.astro-ph/0405087}

\bibitem[{{Dumusque}(2016)}]{Dumusque_2016}
{Dumusque}, X. 2016, \aap, 593, A5, \dodoi{10.1051/0004-6361/201628672}

\bibitem[{{Faria} {et~al.}(2022){Faria}, {Su{\'a}rez Mascare{\~n}o}, {Figueira}, {Silva}, {Damasso}, {Demangeon}, {Pepe}, {Santos}, {Rebolo}, {Cristiani}, {Adibekyan}, {Alibert}, {Allart}, {Barros}, {Cabral}, {D'Odorico}, {Di Marcantonio}, {Dumusque}, {Ehrenreich}, {Gonz{\'a}lez Hern{\'a}ndez}, {Hara}, {Lillo-Box}, {Lo Curto}, {Lovis}, {Martins}, {M{\'e}gevand}, {Mehner}, {Micela}, {Molaro}, {Nunes}, {Pall{\'e}}, {Poretti}, {Sousa}, {Sozzetti}, {Tabernero}, {Udry}, \& {Zapatero Osorio}}]{Faria_2022}
{Faria}, J.~P., {Su{\'a}rez Mascare{\~n}o}, A., {Figueira}, P., {et~al.} 2022, \aap, 658, A115, \dodoi{10.1051/0004-6361/202142337}

\bibitem[{{Gaur} {et~al.}(1973){Gaur}, {Pande}, \& {Tripathi}}]{Gaur_1973}
{Gaur}, V.~P., {Pande}, M.~C., \& {Tripathi}, B.~M. 1973, Bulletin of the Astronomical Institutes of Czechoslovakia, 24, 138

\bibitem[{{Giguere} {et~al.}(2016){Giguere}, {Fischer}, {Zhang}, {Matthews}, {Cameron}, \& {Henry}}]{Giguere_2016}
{Giguere}, M.~J., {Fischer}, D.~A., {Zhang}, C. X.~Y., {et~al.} 2016, \apj, 824, 150, \dodoi{10.3847/0004-637X/824/2/150}

\bibitem[{{Gilbertson} {et~al.}(2020){Gilbertson}, {Ford}, {Jones}, \& {Stenning}}]{Gilbertson_2020}
{Gilbertson}, C., {Ford}, E.~B., {Jones}, D.~E., \& {Stenning}, D.~C. 2020, \apj, 905, 155, \dodoi{10.3847/1538-4357/abc627}

\bibitem[{{Gupta} {et~al.}(2021){Gupta}, {Wright}, {Robertson}, {Halverson}, {Luhn}, {Roy}, {Mahadevan}, {Ford}, {Bender}, {Blake}, {Hearty}, {Kanodia}, {Logsdon}, {McElwain}, {Monson}, {Ninan}, {Schwab}, {Stef{\'a}nsson}, \& {Terrien}}]{Gupta_NEID_2021}
{Gupta}, A.~F., {Wright}, J.~T., {Robertson}, P., {et~al.} 2021, \aj, 161, 130, \dodoi{10.3847/1538-3881/abd79e}

\bibitem[{{Horn} {et~al.}(1997){Horn}, {Staude}, \& {Landgraf}}]{Horn_1997}
{Horn}, T., {Staude}, J., \& {Landgraf}, V. 1997, \solphys, 172, 69, \dodoi{10.1023/A:1004909030878}

\bibitem[{{Howard} {et~al.}(2010){Howard}, {Johnson}, {Marcy}, {Fischer}, {Wright}, {Bernat}, {Henry}, {Peek}, {Isaacson}, {Apps}, {Endl}, {Cochran}, {Valenti}, {Anderson}, \& {Piskunov}}]{Howard_2010}
{Howard}, A.~W., {Johnson}, J.~A., {Marcy}, G.~W., {et~al.} 2010, \apj, 721, 1467, \dodoi{10.1088/0004-637X/721/2/1467}

\bibitem[{{Hu{\'e}lamo} {et~al.}(2008){Hu{\'e}lamo}, {Figueira}, {Bonfils}, {Santos}, {Pepe}, {Gillon}, {Azevedo}, {Barman}, {Fern{\'a}ndez}, {di Folco}, {Guenther}, {Lovis}, {Melo}, {Queloz}, \& {Udry}}]{Huelamo_2008}
{Hu{\'e}lamo}, N., {Figueira}, P., {Bonfils}, X., {et~al.} 2008, \aap, 489, L9, \dodoi{10.1051/0004-6361:200810596}

\bibitem[{{Huerta} {et~al.}(2008){Huerta}, {Johns-Krull}, {Prato}, {Hartigan}, \& {Jaffe}}]{Huerta_2008}
{Huerta}, M., {Johns-Krull}, C.~M., {Prato}, L., {Hartigan}, P., \& {Jaffe}, D.~T. 2008, \apj, 678, 472, \dodoi{10.1086/526415}

\bibitem[{{Jones} {et~al.}(2017){Jones}, {Stenning}, {Ford}, {Wolpert}, {Loredo}, {Gilbertson}, \& {Dumusque}}]{Jones_2017}
{Jones}, D.~E., {Stenning}, D.~C., {Ford}, E.~B., {et~al.} 2017, arXiv e-prints, arXiv:1711.01318, \dodoi{10.48550/arXiv.1711.01318}

\bibitem[{{Kuckein} {et~al.}(2021){Kuckein}, {Balthasar}, {Quintero Noda}, {Diercke}, {Trelles Arjona}, {Ruiz Cobo}, {Felipe}, {Denker}, {Verma}, {Kontogiannis}, \& {Sobotka}}]{Kuckein_2021}
{Kuckein}, C., {Balthasar}, H., {Quintero Noda}, C., {et~al.} 2021, \aap, 653, A165, \dodoi{10.1051/0004-6361/202140596}

\bibitem[{{Lafarga} {et~al.}(2020){Lafarga}, {Ribas}, {Lovis}, {Perger}, {Zechmeister}, {Bauer}, {K{\"u}rster}, {Cort{\'e}s-Contreras}, {Morales}, {Herrero}, {Rosich}, {Baroch}, {Reiners}, {Caballero}, {Quirrenbach}, {Amado}, {Alacid}, {B{\'e}jar}, {Dreizler}, {Hatzes}, {Henning}, {Jeffers}, {Kaminski}, {Montes}, {Pedraz}, {Rodr{\'\i}guez-L{\'o}pez}, \& {Schmitt}}]{Lafarga_2020}
{Lafarga}, M., {Ribas}, I., {Lovis}, C., {et~al.} 2020, \aap, 636, A36, \dodoi{10.1051/0004-6361/201937222}

\bibitem[{{Lafarga} {et~al.}(2023){Lafarga}, {Ribas}, {Zechmeister}, {Reiners}, {L{\'o}pez-Gallifa}, {Montes}, {Quirrenbach}, {Amado}, {Caballero}, {Azzaro}, {B{\'e}jar}, {Hatzes}, {Henning}, {Jeffers}, {Kaminski}, {K{\"u}rster}, {Sch{\"o}fer}, {Schweitzer}, {Tabernero}, \& {Osorio}}]{Lafarga_2023}
{Lafarga}, M., {Ribas}, I., {Zechmeister}, M., {et~al.} 2023, \aap, 674, A61, \dodoi{10.1051/0004-6361/202245602}

\bibitem[{{Landstreet}(2015)}]{Landstreet_2015}
{Landstreet}, J.~D. 2015, in New Windows on Massive Stars, ed. G.~{Meynet}, C.~{Georgy}, J.~{Groh}, \& P.~{Stee}, Vol. 307, 311--320, \dodoi{10.1017/S1743921314007017}

\bibitem[{{Lee} {et~al.}(2023){Lee}, {Jeong}, {Koo}, {Lim}, {Park}, {Bang}, {Choi}, {Oh}, \& {Han}}]{Lee_2023}
{Lee}, B.-C., {Jeong}, G., {Koo}, J.-R., {et~al.} 2023, Journal of Korean Astronomical Society, 56, 195, \dodoi{10.5303/JKAS.2023.56.2.195}

\bibitem[{{Li} {et~al.}(2024){Li}, {Kane}, {Brandt}, {Fetherolf}, {Robertson}, {Zhao}, {Dalba}, {Wittenmyer}, {Butler}, {Diaz}, {Howell}, {Bailey}, {Carter}, {Furlan}, {Gnilka}, {Jones}, {O'Toole}, \& {Tinney}}]{Li_2024}
{Li}, Z., {Kane}, S.~R., {Brandt}, T.~D., {et~al.} 2024, arXiv e-prints, arXiv:2401.17415, \dodoi{10.48550/arXiv.2401.17415}

\bibitem[{{Lites} {et~al.}(1998){Lites}, {Thomas}, {Bogdan}, \& {Cally}}]{Lites_1998}
{Lites}, B.~W., {Thomas}, J.~H., {Bogdan}, T.~J., \& {Cally}, P.~S. 1998, \apj, 497, 464, \dodoi{10.1086/305451}

\bibitem[{{Livingston} \& {Watson}(2015)}]{Livingston_2015}
{Livingston}, W., \& {Watson}, F. 2015, \grl, 42, 9185, \dodoi{10.1002/2015GL065413}

\bibitem[{{Lozitsky}(2016)}]{Lozitsky_2016}
{Lozitsky}, V.~G. 2016, Advances in Space Research, 57, 398, \dodoi{10.1016/j.asr.2015.08.032}

\bibitem[{{Lozitsky} {et~al.}(2020){Lozitsky}, {Osipov}, \& {Stodilka}}]{Lozitsky_2020}
{Lozitsky}, V.~G., {Osipov}, S.~M., \& {Stodilka}, M.~I. 2020, Odessa Astronomical Publications, 33, 89, \dodoi{10.18524/1810-4215.2020.33.216451}

\bibitem[{{Mathur} {et~al.}(2024){Mathur}, {Nagaraju}, {Yadav}, \& {Joshi}}]{Mathur_2024}
{Mathur}, H., {Nagaraju}, K., {Yadav}, R., \& {Joshi}, J. 2024, arXiv e-prints, arXiv:2406.02083, \dodoi{10.48550/arXiv.2406.02083}

\bibitem[{{Mayor} \& {Queloz}(1995)}]{Mayor_1995}
{Mayor}, M., \& {Queloz}, D. 1995, \nat, 378, 355, \dodoi{10.1038/378355a0}

\bibitem[{{Meunier} \& {Lagrange}(2013)}]{Meunier_Lagrange_2013}
{Meunier}, N., \& {Lagrange}, A.~M. 2013, \aap, 551, A101, \dodoi{10.1051/0004-6361/201219917}

\bibitem[{{Moore} {et~al.}(1966){Moore}, {Minnaert}, \& {Houtgast}}]{Moore_1966}
{Moore}, C.~E., {Minnaert}, M.~G.~J., \& {Houtgast}, J. 1966, {The solar spectrum 2935 A to 8770 A} (Washington: US Government Printing Office)

\bibitem[{{Morris} {et~al.}(2019){Morris}, {Curtis}, {Sakari}, {Hawley}, \& {Agol}}]{Morris_2019}
{Morris}, B.~M., {Curtis}, J.~L., {Sakari}, C., {Hawley}, S.~L., \& {Agol}, E. 2019, \aj, 158, 101, \dodoi{10.3847/1538-3881/ab2e04}

\bibitem[{Ozerov \& Vorobyev(2007)}]{Ruslan_2007}
Ozerov, R.~P., \& Vorobyev, A.~A. 2007, in Physics for Chemists, ed. R.~P. Ozerov \& A.~A. Vorobyev (Amsterdam: Elsevier), 423--496, \dodoi{https://doi.org/10.1016/B978-044452830-8/50009-X}

\bibitem[{{Pande} \& {Gaur}(1975)}]{Pande_1975}
{Pande}, M.~C., \& {Gaur}, V.~P. 1975, \nat, 253, 104, \dodoi{10.1038/253104a0}

\bibitem[{{Pepe} {et~al.}(2021){Pepe}, {Cristiani}, {Rebolo}, {Santos}, {Dekker}, {Cabral}, {Di Marcantonio}, {Figueira}, {Lo Curto}, {Lovis}, {Mayor}, {M{\'e}gevand}, {Molaro}, {Riva}, {Zapatero Osorio}, {Amate}, {Manescau}, {Pasquini}, {Zerbi}, {Adibekyan}, {Abreu}, {Affolter}, {Alibert}, {Aliverti}, {Allart}, {Allende Prieto}, {{\'A}lvarez}, {Alves}, {Avila}, {Baldini}, {Bandy}, {Barros}, {Benz}, {Bianco}, {Borsa}, {Bourrier}, {Bouchy}, {Broeg}, {Calderone}, {Cirami}, {Coelho}, {Conconi}, {Coretti}, {Cumani}, {Cupani}, {D'Odorico}, {Damasso}, {Deiries}, {Delabre}, {Demangeon}, {Dumusque}, {Ehrenreich}, {Faria}, {Fragoso}, {Genolet}, {Genoni}, {G{\'e}nova Santos}, {Gonz{\'a}lez Hern{\'a}ndez}, {Hughes}, {Iwert}, {Kerber}, {Knudstrup}, {Landoni}, {Lavie}, {Lillo-Box}, {Lizon}, {Maire}, {Martins}, {Mehner}, {Micela}, {Modigliani}, {Monteiro}, {Monteiro}, {Moschetti}, {Murphy}, {Nunes}, {Oggioni}, {Oliveira}, {Oshagh}, {Pall{\'e}}, {Pariani}, {Poretti}, {Rasilla}, {Rebord{\~a}o}, {Redaelli}, {Santana Tschudi},
  {Santin}, {Santos}, {S{\'e}gransan}, {Schmidt}, {Segovia}, {Sosnowska}, {Sozzetti}, {Sousa}, {Span{\`o}}, {Su{\'a}rez Mascare{\~n}o}, {Tabernero}, {Tenegi}, {Udry}, \& {Zanutta}}]{Pepe_ESPRESSO_2021}
{Pepe}, F., {Cristiani}, S., {Rebolo}, R., {et~al.} 2021, \aap, 645, A96, \dodoi{10.1051/0004-6361/202038306}

\bibitem[{{Petersburg} {et~al.}(2020){Petersburg}, {Ong}, {Zhao}, {Blackman}, {Brewer}, {Buchhave}, {Cabot}, {Davis}, {Jurgenson}, {Leet}, {McCracken}, {Sawyer}, {Sharov}, {Tronsgaard}, {Szymkowiak}, \& {Fischer}}]{Petersburg_EXPRES_2020}
{Petersburg}, R.~R., {Ong}, J.~M.~J., {Zhao}, L.~L., {et~al.} 2020, \aj, 159, 187, \dodoi{10.3847/1538-3881/ab7e31}

\bibitem[{{Pevtsov} {et~al.}(2021){Pevtsov}, {Bertello}, {Nagovitsyn}, {Tlatov}, \& {Pipin}}]{Pevtsov_2021}
{Pevtsov}, A.~A., {Bertello}, L., {Nagovitsyn}, Y.~A., {Tlatov}, A.~G., \& {Pipin}, V.~V. 2021, Journal of Space Weather and Space Climate, 11, 4, \dodoi{10.1051/swsc/2020069}

\bibitem[{{Przybilla} {et~al.}(2011){Przybilla}, {Nieva}, \& {Butler}}]{Przybilla_2011}
{Przybilla}, N., {Nieva}, M.-F., \& {Butler}, K. 2011, in Journal of Physics Conference Series, Vol. 328, Journal of Physics Conference Series, 012015, \dodoi{10.1088/1742-6596/328/1/012015}

\bibitem[{{Rajpaul} {et~al.}(2020){Rajpaul}, {Aigrain}, \& {Buchhave}}]{Rajpaul_2020}
{Rajpaul}, V.~M., {Aigrain}, S., \& {Buchhave}, L.~A. 2020, \mnras, 492, 3960, \dodoi{10.1093/mnras/stz3599}

\bibitem[{{Reiners}(2012)}]{Reiners_2012}
{Reiners}, A. 2012, Living Reviews in Solar Physics, 9, 1, \dodoi{10.12942/lrsp-2012-1}

\bibitem[{{Robertson} {et~al.}(2014){Robertson}, {Mahadevan}, {Endl}, \& {Roy}}]{Robertson_2014}
{Robertson}, P., {Mahadevan}, S., {Endl}, M., \& {Roy}, A. 2014, Science, 345, 440, \dodoi{10.1126/science.1253253}

\bibitem[{{Robertson} {et~al.}(2020){Robertson}, {Stefansson}, {Mahadevan}, {Endl}, {Cochran}, {Beard}, {Bender}, {Diddams}, {Duong}, {Ford}, {Fredrick}, {Halverson}, {Hearty}, {Holcomb}, {Juan}, {Kanodia}, {Lubin}, {Metcalf}, {Monson}, {Ninan}, {Palafoutas}, {Ramsey}, {Roy}, {Schwab}, {Terrien}, \& {Wright}}]{Robertson_2020}
{Robertson}, P., {Stefansson}, G., {Mahadevan}, S., {et~al.} 2020, \apj, 897, 125, \dodoi{10.3847/1538-4357/ab989f}

\bibitem[{{Rothman} {et~al.}(2009){Rothman}, {Gordon}, {Barbe}, {Benner}, {Bernath}, {Birk}, {Boudon}, {Brown}, {Campargue}, {Champion}, {Chance}, {Coudert}, {Dana}, {Devi}, {Fally}, {Flaud}, {Gamache}, {Goldman}, {Jacquemart}, {Kleiner}, {Lacome}, {Lafferty}, {Mandin}, {Massie}, {Mikhailenko}, {Miller}, {Moazzen-Ahmadi}, {Naumenko}, {Nikitin}, {Orphal}, {Perevalov}, {Perrin}, {Predoi-Cross}, {Rinsland}, {Rotger}, {{\v{S}}ime{\v{c}}kov{\'a}}, {Smith}, {Sung}, {Tashkun}, {Tennyson}, {Toth}, {Vandaele}, \& {Vander Auwera}}]{Rothman_2009}
{Rothman}, L.~S., {Gordon}, I.~E., {Barbe}, A., {et~al.} 2009, \jqsrt, 110, 533, \dodoi{10.1016/j.jqsrt.2009.02.013}

\bibitem[{{Rutten}(1988)}]{Rutten_1988}
{Rutten}, R.~J. 1988, in Astrophysics and Space Science Library, Vol. 138, IAU Colloq. 94: Physics of Formation of FE II Lines Outside LTE, ed. R.~{Viotti}, A.~{Vittone}, \& M.~{Friedjung}, 185--210, \dodoi{10.1007/978-94-009-4023-9_23}

\bibitem[{{Rutten} \& {Kostik}(1982)}]{Rutten_1982}
{Rutten}, R.~J., \& {Kostik}, R.~I. 1982, \aap, 115, 104

\bibitem[{{Ryabchikova} {et~al.}(2015){Ryabchikova}, {Piskunov}, {Kurucz}, {Stempels}, {Heiter}, {Pakhomov}, \& {Barklem}}]{Ryabchikova_2015}
{Ryabchikova}, T., {Piskunov}, N., {Kurucz}, R.~L., {et~al.} 2015, \physscr, 90, 054005, \dodoi{10.1088/0031-8949/90/5/054005}

\bibitem[{{Saar} {et~al.}(1998){Saar}, {Butler}, \& {Marcy}}]{Saar_1998}
{Saar}, S.~H., {Butler}, R.~P., \& {Marcy}, G.~W. 1998, \apjl, 498, L153, \dodoi{10.1086/311325}

\bibitem[{{Saar} \& {Donahue}(1997)}]{Saar_Donahue_1997}
{Saar}, S.~H., \& {Donahue}, R.~A. 1997, \apj, 485, 319, \dodoi{10.1086/304392}

\bibitem[{{Saar} \& {Fischer}(2000)}]{Saar_Fischer_2000}
{Saar}, S.~H., \& {Fischer}, D. 2000, \apjl, 534, L105, \dodoi{10.1086/312648}

\bibitem[{Saptari(2003)}]{saptari2003fourier}
Saptari, V. 2003, Fourier transform spectroscopy instrumentation engineering (SPIE Optical Engineering Press Bellingham Washington, DC)

\bibitem[{{Seifahrt} {et~al.}(2022){Seifahrt}, {Bean}, {Kasper}, {St{\"u}rmer}, {Brady}, {Liu}, {Zechmeister}, {Stef{\'a}nsson}, {Montet}, {White}, {Tapia}, {Mocnik}, {Xu}, \& {Schwab}}]{Seifahrt_MAROONX_2022}
{Seifahrt}, A., {Bean}, J.~L., {Kasper}, D., {et~al.} 2022, in Society of Photo-Optical Instrumentation Engineers (SPIE) Conference Series, Vol. 12184, Ground-based and Airborne Instrumentation for Astronomy IX, ed. C.~J. {Evans}, J.~J. {Bryant}, \& K.~{Motohara}, 121841G, \dodoi{10.1117/12.2629428}

\bibitem[{{Siu-Tapia} {et~al.}(2017){Siu-Tapia}, {Lagg}, {Solanki}, {van Noort}, \& {Jur{\v{c}}{\'a}k}}]{SiuTapia_2017}
{Siu-Tapia}, A., {Lagg}, A., {Solanki}, S.~K., {van Noort}, M., \& {Jur{\v{c}}{\'a}k}, J. 2017, \aap, 607, A36, \dodoi{10.1051/0004-6361/201730647}

\bibitem[{{Skelly} {et~al.}(2008){Skelly}, {Unruh}, {Collier Cameron}, {Barnes}, {Donati}, {Lawson}, \& {Carter}}]{Skelly_2008}
{Skelly}, M.~B., {Unruh}, Y.~C., {Collier Cameron}, A., {et~al.} 2008, \mnras, 385, 708, \dodoi{10.1111/j.1365-2966.2008.12917.x}

\bibitem[{{Smitha} {et~al.}(2020){Smitha}, {Holzreuter}, {van Noort}, \& {Solanki}}]{Smitha_2020}
{Smitha}, H.~N., {Holzreuter}, R., {van Noort}, M., \& {Solanki}, S.~K. 2020, \aap, 633, A157, \dodoi{10.1051/0004-6361/201937041}

\bibitem[{{Smitha} {et~al.}(2021){Smitha}, {Holzreuter}, {van Noort}, \& {Solanki}}]{Smitha_2021}
---. 2021, \aap, 647, A46, \dodoi{10.1051/0004-6361/202039107}

\bibitem[{{Smitha} {et~al.}(2023){Smitha}, {van Noort}, {Solanki}, \& {Castellanos Dur{\'a}n}}]{Smitha_2023}
{Smitha}, H.~N., {van Noort}, M., {Solanki}, S.~K., \& {Castellanos Dur{\'a}n}, J.~S. 2023, \aap, 669, A144, \dodoi{10.1051/0004-6361/202245130}

\bibitem[{{Solanki}(2003)}]{Solanki_2003}
{Solanki}, S.~K. 2003, \aapr, 11, 153, \dodoi{10.1007/s00159-003-0018-4}

\bibitem[{{Solanki} \& {Steenbock}(1988)}]{Solanki_1988}
{Solanki}, S.~K., \& {Steenbock}, W. 1988, \aap, 189, 243

\bibitem[{{Stift} \& {Leone}(2003)}]{Stift_2003}
{Stift}, M.~J., \& {Leone}, F. 2003, \aap, 398, 411, \dodoi{10.1051/0004-6361:20021605}

\bibitem[{{Tennyson} {et~al.}(2024){Tennyson}, {Yurchenko}, {Zhang}, {Bowesman}, {Brady}, {Buldyreva}, {Chubb}, {Gamache}, {Gorman}, {Guest}, {Hill}, {Kefala}, {Lynas-Gray}, {Mellor}, {McKemmish}, {Mitev}, {Mizus}, {Owens}, {Peng}, {Perri}, {Pezzella}, {Polyansky}, {Qu}, {Semenov}, {Smola}, {Solokov}, {Somogyi}, {Upadhyay}, {Wright}, \& {Zobov}}]{Tennyson_2024}
{Tennyson}, J., {Yurchenko}, S.~N., {Zhang}, J., {et~al.} 2024, \jqsrt, 326, 109083, \dodoi{10.1016/j.jqsrt.2024.109083}

\bibitem[{Thomas \& Weiss(1992)}]{Thomas_1992}
Thomas, J.~H., \& Weiss, N.~O., eds. 1992, NATO Advanced Study Institute (ASI) Series C, Vol. 375, {Sunspots - Theory \& Observations: NATO Cambridge, 1992}, \dodoi{10.1007/978-94-011-2769-1}

\bibitem[{{Valenti} \& {Piskunov}(1996)}]{Valenti_Piskunov_1996}
{Valenti}, J.~A., \& {Piskunov}, N. 1996, \aaps, 118, 595

\bibitem[{{Vanderburg} {et~al.}(2016){Vanderburg}, {Plavchan}, {Johnson}, {Ciardi}, {Swift}, \& {Kane}}]{Vanderburg_2016}
{Vanderburg}, A., {Plavchan}, P., {Johnson}, J.~A., {et~al.} 2016, \mnras, 459, 3565, \dodoi{10.1093/mnras/stw863}

\bibitem[{{Vogt} {et~al.}(1994){Vogt}, {Allen}, {Bigelow}, {Bresee}, {Brown}, {Cantrall}, {Conrad}, {Couture}, {Delaney}, {Epps}, {Hilyard}, {Hilyard}, {Horn}, {Jern}, {Kanto}, {Keane}, {Kibrick}, {Lewis}, {Osborne}, {Pardeilhan}, {Pfister}, {Ricketts}, {Robinson}, {Stover}, {Tucker}, {Ward}, \& {Wei}}]{Vogt_1994}
{Vogt}, S.~S., {Allen}, S.~L., {Bigelow}, B.~C., {et~al.} 1994, in Society of Photo-Optical Instrumentation Engineers (SPIE) Conference Series, Vol. 2198, Instrumentation in Astronomy VIII, ed. D.~L. {Crawford} \& E.~R. {Craine}, 362, \dodoi{10.1117/12.176725}

\bibitem[{{von Stauffenberg} {et~al.}(2024){von Stauffenberg}, {Trifonov}, {Quirrenbach}, {Reffert}, {Kaminski}, {Dreizler}, {Ribas}, {Reiners}, {K{\"u}rster}, {Twicken}, {Rapetti}, {Caballero}, {Amado}, {B{\'e}jar}, {Cifuentes}, {G{\'o}ngora}, {Hatzes}, {Henning}, {Montes}, {Morales}, \& {Schweitzer}}]{vonStauffenberg_2024}
{von Stauffenberg}, A., {Trifonov}, T., {Quirrenbach}, A., {et~al.} 2024, \aap, 688, A112, \dodoi{10.1051/0004-6361/202449375}

\bibitem[{{Wallace} {et~al.}(2005){Wallace}, {Hinkle}, \& {Livingston}}]{Wallace_2005}
{Wallace}, L., {Hinkle}, K., \& {Livingston}, W.~C. 2005, {An atlas of sunspot umbral spectra in the visible from 15,000 to 25,500 cm-{\textonesuperior} (3920 to 6664 {\r{A}})} (NSO)

\bibitem[{{Wallace} {et~al.}(2011){Wallace}, {Hinkle}, {Livingston}, \& {Davis}}]{Wallace_2011}
{Wallace}, L., {Hinkle}, K.~H., {Livingston}, W.~C., \& {Davis}, S.~P. 2011, \apjs, 195, 6, \dodoi{10.1088/0067-0049/195/1/6}

\bibitem[{{Wilson} \& {Casey}(2023)}]{Wilson_2023}
{Wilson}, T.~A., \& {Casey}, A.~R. 2023, \mnras, 524, 731, \dodoi{10.1093/mnras/stad1875}

\bibitem[{{Zhao} {et~al.}(2022){Zhao}, {Fischer}, {Ford}, {Wise}, {Cretignier}, {Aigrain}, {Barragan}, {Bedell}, {Buchhave}, {Camacho}, {Cegla}, {Cisewski-Kehe}, {Collier Cameron}, {de Beurs}, {Dodson-Robinson}, {Dumusque}, {Faria}, {Gilbertson}, {Haley}, {Harrell}, {Hogg}, {Holzer}, {John}, {Klein}, {Lafarga}, {Lienhard}, {Maguire-Rajpaul}, {Mortier}, {Nicholson}, {Palumbo}, {Ramirez Delgado}, {Shallue}, {Vanderburg}, {Viana}, {Zhao}, {Zicher}, {Cabot}, {Henry}, {Roettenbacher}, {Brewer}, {Llama}, {Petersburg}, \& {Szymkowiak}}]{Zhao_2022}
{Zhao}, L.~L., {Fischer}, D.~A., {Ford}, E.~B., {et~al.} 2022, \aj, 163, 171, \dodoi{10.3847/1538-3881/ac5176}

\end{thebibliography}
\bibliographystyle{aasjournal}

\end{document}